\documentstyle[12pt]{article}
\catcode`\@=11
\def\maketitle{\par
 \begingroup
 \def\thefootnote{\fnsymbol{footnote}}
 \def\@makefnmark{\hbox
 to 0pt{$^{\@thefnmark}$\hss}}
 \if@twocolumn
 \twocolumn[\@maketitle]
 \else \newpage
 \global\@topnum\z@ \@maketitle \fi\thispagestyle{empty}\@thanks
 \endgroup
 \setcounter{footnote}{0}
 \let\maketitle\relax
 \let\@maketitle\relax
 \gdef\@thanks{}\gdef\@author{}\gdef\@title{}\let\thanks\relax}
\def\@maketitle{\newpage
 \null
 \hbox to\textwidth{\hfil\hbox{\begin{tabular}{r}\@preprint\end{tabular}}}
 \vskip 2em \begin{center}
 {\Large\bf \@title \par} \vskip 1.5em {\normalsize \lineskip .5em
\begin{tabular}[t]{c}\@author
 \end{tabular}\par}
 \end{center}
 \par
 \vskip 1.5em}
\def\preprint#1{\gdef\@preprint{#1}}
\def\abstract{\if@twocolumn
\section*{Abstract}
\else \normalsize
\begin{center}
{\large\bf Abstract\vspace{-.5em}\vspace{0pt}}
\end{center}
\quotation
\fi}
\def\endabstract{\if@twocolumn\else\endquotation\fi}
\catcode`\@=12

\begin{document}
\baselineskip=.285in

\preprint{SNUTP-96-010}

\title{\Large\bf Lattice Gauge Theory \\ of\\
Three Dimensional Thirring Model
\protect\\[1mm]\ }
\author{\normalsize Seyong Kim\\[2mm]
{\normalsize\it Center for Theoretical Physics, Seoul National University,
Seoul 151-742, Korea}\\[2mm]
{\normalsize\it skim$@$ctp.snu.ac.kr}\\[6mm]
{\normalsize Yoonbai Kim}\\[2mm]
{\normalsize\it Department of Physics, Pusan National University, 
Pusan 609-735, Korea}\\[2mm]
{\normalsize\it yoonbai$@$top.phys.pusan.ac.kr}} 
\date{}
\maketitle

\begin{center}
{\large\bf Abstract}\\[3mm]
\end{center}
\indent\indent Three dimensional Thirring model with $N$ four-component 
Dirac fermions, reformulated as a lattice gauge theory, is studied 
by computer simulation. 
According to an $8^{3}$ data and preliminary $16^3$ data, chiral symmetry 
is found to be spontaneously broken for $N=2,\;4$ and 6.
$N=2$ data exhibits long tail of 
the non-vanishing chiral condensate into weak coupling region, and $N=6$ 
case shows phase separation between the strong coupling region and 
the weak coupling region. 
Although the comparison between $8^3$ data and $16^3$ data shows large
finite volume effects, an existence of the critical fermion flavor number 
$N_{{\rm cr}}$ $(2<N_{{\rm cr}}<6)$ for which the chiral behavior 
changes its character, is suggested by the current numerical data.

\vspace{2mm}

\noindent PACS number(s): 11.15.Ha, 11.30.Qc, 11.30.Rd

\newpage

\pagenumbering{arabic}
\thispagestyle{plain}

Thirring model, appeared in the (1+1) dimensional world around forty years 
ago, has shed light on understanding various intriguing issues of field
theories, {\it e.g.}, exactly-soluble models, bosonization, conformal
fixed point, etc. Recently, the very model with $N$ four-component Dirac
fermions has attracted attention in the (2+1) dimensional world 
with regard to renormalizability of four-fermion models
in the context of $1/N$-expansion \cite{Par}-\cite{Han}, dynamical mass 
generation for the fermion in the analysis of Schwinger-Dyson equation
\cite{GMRS}-\cite{Kon} and so on. In the listed works, usually $1/N$ expansion
technique is employed since interesting features of the model are 
inherently non-perturbative. In this note, using a Monte Carlo simulation 
technique, we consider the Thirring model on 
a three-dimensional(3D) cubic lattice, and ask whether the vector-type four 
fermion interaction of the model breaks chiral symmetry spontaneously.

The Lagrangian for the 3D Thirring model in 
Euclidean space time is given by 
\begin{equation}
{\cal L}=\sum_{a}\bar{\psi}_{a}(\gamma_{\mu}\partial_{\mu}+m)\psi_{a} 
+\frac{G}{2N}(\sum_{a}\bar{\psi}_{a}\gamma_{\mu}\psi_{a})^{2} .
\end{equation}
Typical way of treating this model is to introduce an auxiliary vector field
$A_{\mu}$ \cite{Par,Han,GMRS,HP,RS}. On a lattice, this leads to an
auxiliary vector field formulation \cite{DH}. 
An alternative approach which is attractive to us, is to regard $A_{\mu}$ 
as a $U(1)$ gauge field. Indeed a reformulation of the Thirring 
model as a gauge theory can be achieved through a hidden local 
symmetry \cite{BKY}, \cite{KK,IKSY,Kon}: 
\begin{equation}\label{HLSac}
{\cal L}_{\rm HLS}=\sum_{a}\bar{\psi}_{a}(\gamma_{\mu}D_{\mu}+m)
\psi_{a}+\frac{1}{2G}(A_{\mu}-\sqrt{N}\partial_{\mu}\phi)^{2},
\end{equation}
where $\psi_{a}$ is a 4-component Dirac spinor, $\phi$ is a fictitious
Nambu-Goldstone boson field (or equivalently a St\"{u}ckelberg field), 
and $D_\mu = \partial_\mu - \frac{i}{\sqrt{N}} A_\mu$. 

Noticing that the unitary gauge-fixed form of the 
above $U(1)$ gauge theory coincides exactly with the original Thirring model, 
we expect benefits from gauge-invariant formulation. 
One such benefit is that, in the limit of zero bare fermion mass, 
one can avoid complications like the dynamical generation of 
parity-violating mass \cite{VW} or Chern-Simons term \cite{Red}.  
The other is that we have a freedom to choose the most appropriate gauge for
each particular situations of interest, {\it e.g.}, analysis of the 
Schwinger-Dyson equation in a rigorous fashion \cite{IKSY} and a 
construction of a gauge-invariant effective potential for the chiral 
order parameter \cite{Kon}.  

Since the model we are considering is free from parity violation
by an exact argument in the gauge theory formulation, we can focus on 
the spontaneous breakdown of 
chiral symmetry (actually a flavor symmetry breaking 
of $U(2N)\rightarrow U(N)\times U(N)$). There are various interesting aspects
to be addressed once the above chiral symmetry breaking occurs, 
which were raised by analytic methods. First, a rather systematic 
approach of Schwinger-Dyson equations in the gauge theory 
formulation \cite{IKSY} suggests a finite, critical number of 
the four-component fermion species,
\begin{equation}
N_{\rm cr}=\frac{128}{3\pi^{2}}\approx 4.32\; 
\end{equation} 
which is obtained in the infinite limit of four-fermion coupling $g$ 
($g$ is a dimensionless coupling, $G$ rescaled by the ultraviolet 
cutoff $\Lambda, g\equiv G\Lambda$).
In contrast, the auxiliary field formulation with the ``{\it Landau 
gauge fixing}'' implies otherwise.\footnote{In fact this Landau gauge fixing
amounts to an $R_{\xi}$ gauge in the hidden local symmetry formulation
if we concede that the auxiliary field is a gauge field.} 
A crude constant mass approximation to this formulation seems 
to predict a chiral symmetry breaking for all $N$ \cite{HP} 
\begin{equation}
g_{\rm cr}\propto\exp\Bigl(\frac{\pi^{2}}{16}N\Bigr),
\end{equation}
which also tells us the ratio of the values of the critical coupling 
$g_{\rm cr}$ for different $N$'s. Until now, there is no known equivalent 
relationship between $g_{\rm cr}$ and $N$ in the gauge theory formulation. 
However, if such a relation exists, it should be different from the above
equation since there exists a critical number of fermion flavor $N_{\rm cr}$
in the gauge theory formulation \cite{SSY}.
Second, it may be interesting if we can read the chiral condensate
$\bigl<\bar{\psi}\psi\bigr>$ as a function of the coupling $g$ for
a given $N$. In the gauge theory formulation \cite{Kon}, a critical scaling
behavior was suggested to follow
\begin{equation}\label{psib}
\bigl<\bar{\psi}\psi\bigr>\approx A_{N}\exp\biggl(
-\frac{B_{N}}{\sqrt{\frac{g}{g_{\rm cr}}-1}}\biggr).
\end{equation} 
Of course, on a finite lattice, the behavior of the chiral condensate will 
be modified by finite volume effect and may be difficult to compare with 
this formula away from $g_{\rm cr}$.

Analytic studies in the framework of the Schwinger-Dyson
equations combined with an $1/N$ expansion raised the questions in the above
and it is interesting to investigate these questions by a lattice study. 
In choosing an appropriate action for numerical simulation, one can use 
an auxiliary vector field formulation without gauge symmetry or employ a gauge
field formulation based on a hidden local symmetry. 
Recently simulation results of the auxiliary field formulation case was 
reported in Ref.\cite{DH}. Here we adopt the gauge theory formulation and take
advantage of various benefits coming from the gauge theory formulation.
A discretized version of the 3D action, ${\cal L}_{L}$
with a staggered fermion formulation for the fermion part, is
\begin{equation}\label{latac}
{\cal L}_{L}=\sum_{a}{\Phi^{\dagger}}_{a}(M^{\dagger}M)^{-1}\Phi_{a}
-N\beta\sum_{\mu}\cos{(\phi(x+\mu)-\phi(x)+\theta_{\mu}(x))},
\end{equation}
where \(\Phi\) is the pseudo fermion and \(\theta_{\mu}\) is the gauge field, 
\begin{equation}\label{mxy}
M_{x,y}=\frac{1}{2}\sum_{\mu}\eta_{\mu}(x)(e^{i\theta_{\mu}(x)}\delta_{y,x+\mu}
-e^{-i\theta_{\mu}(x-\mu)}\delta_{y,x-\mu})+m\delta_{x,y} 
\end{equation}
with staggered fermion phase factor, $\eta_\mu (x)$, and 
$\beta = 1/g$, after dropping the constant piece. We include a finite
bare fermion mass term and investigate the massless limit by 
simulating successively smaller fermion mass, $m$, since our numerical method
prohibits simulating the chiral limit directly. The continuum limit
of this lattice model is free from parity-violating mass or Chern-Simons 
term \cite{BB}. The gauge theory formulation of 3D Thirring model 
in Eq.(\ref{latac}) looks like 3-dimensional quantum 
electrodynamics (QED) + scalar QED with 
kinetic term of the gauge field and the Higgs degrees of freedom truncated. 
Note that the link variable $\exp(i\theta_{\mu})$ can not be used for the 
vector-fermion interaction term, $\bar{\psi}A_{\mu}\gamma_{\mu}\psi$, 
in the auxiliary field formulation. This was not encountered in lattice QED3 
case because the gauge symmetry fixes uniquely the interaction between 
the fermion and the gauge field and it is the same as Eq.(\ref{mxy}).
Since this theory is non-asymptotically free, such differences may lead
to different continuum limits. In addition, since in our gauge field 
formulation the gauge kinetic term will be induced
by quantum effects, we can not say whether it will follow compact or 
non-compact form yet. Such an effect seems to be realized indeed by 
some kind of
discontinuous behavior in the actual simulations for the strong coupling 
region \cite{DH}. 

Thus far, we have completed simulations for $N=2,\;4$ and 6 systems on a 
$8^{3}$ lattice volume with fermion mass, $m=0.05,\;0.025$, and 0.0125 by use
of hybrid Monte Carlo algorithm (HMC). Preliminary data on $16^3$ lattice
for $N = 2$ and 6 will be used for finite volume effect study.
As usual, by defining fermion fields 
on even sites only, one can avoid an additional fermion doubling 
from $M^{\dagger}M$ and can simulate even $N$ system with this HMC 
algorithm \cite{DKK}. For odd $N$'s, we need to use other numerical algorithm
such as hybrid molecular dynamics method \cite{Kim_Kim}.
Typically, we have accumulated \(\tau\sim 10,000\) time unit after discarding
initial \(\tau = 100\) time unit as the equilibration time with 
choosing appropriate $\Delta\tau$($= 0.025, 0.02, 0.0125$) which gives
reasonable acceptance rate (50\% to 90\%). The Metropolis step
interval is set to $\tau=0.5$.
As an order parameter for the chiral symmetry, we calculate the chiral 
condensate 
\begin{equation}
\bigl<\bar{\psi}\psi\bigr> = \bigl< {\rm Tr} M^{-1} \bigr>,
\end{equation}
where we use a single noise vector to estimate the condensate.
In addition, we also calculate internal energy $\bigl<E_\beta\bigr>$ defined 
by $\bigl<N\beta (3 - \sum_{\mu}\cos{(\phi(x+\mu)-\phi(x)+\theta_{\mu}(x))})
\bigr>$,
and specific heat $\bigl<{E_\beta}^2\bigr> - \bigl<E_\beta\bigr>^2$ for 
each $N$. Mass spectra for the fermion, \(\pi\)(pseudo scalar channel) and
\(\rho\)(vector channel) are also calculated. However, due to a short 
extent in the time direction, we were not able to fit the propagators 
reliably. Thus, in the following discussion, we will rely mainly on 
the chiral condensate measurement. 

Qualitative nature of our numerical simulation result can be summarized in 
Fig. 1. The figure is the $m \rightarrow 0$ extrapolated chiral 
condensate on a $8^3$ lattice volume for $N=2,\;4$, and 6 respectively 
as a function of $\beta$,
where we assumed a naive linear behavior in $m$ for the chiral condensate
($\sim c_0 + c_1 m$). 
The figure is similar to QED3 case;
in \cite{DKK}, $N = 2$ shows smooth change of the chiral condensate and a 
long tail into weak coupling regime, $N = 3$ shows a complex
behavior of the condensate near the critical point, and $N = 5$ 
shows an abrupt change of the condensate. This supports
the existence of the critical number of fermion flavors for chiral
symmetry breaking in QED3 \cite{ANW}. Likewise, if we believe that the 
naive linear extrapolation of the condensate is a correct procedure
in the gauge theory formulation of 3D Thirring model, 
there appears 
to be a critical $N$ around which the phase transition 
changes its character. A complicated second order phase transition
which connects the strong coupling region to the weak coupling region
smoothly ($N=2$) turns into a phase transition 
which has a phase separation ($N=6$). 

\vspace{4mm}

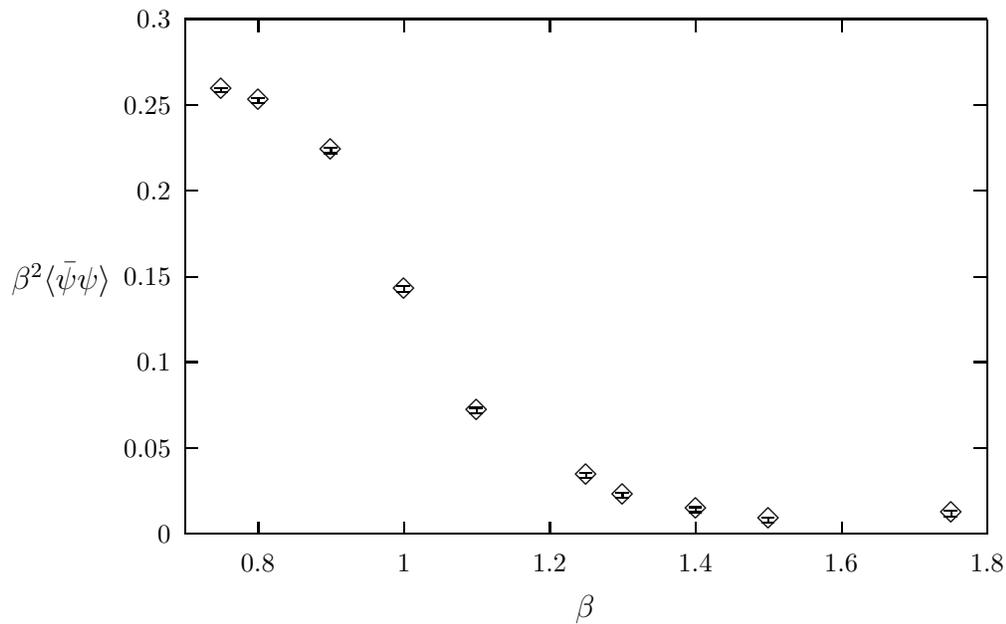
\setlength{\unitlength}{0.240900pt}
\ifx\plotpoint\undefined\newsavebox{\plotpoint}\fi
\sbox{\plotpoint}{\rule[-0.200pt]{0.400pt}{0.400pt}}%
\begin{picture}(1500,900)(0,0)
\font\gnuplot=cmr10 at 10pt
\gnuplot
\sbox{\plotpoint}{\rule[-0.200pt]{0.400pt}{0.400pt}}%
\put(70,468){\makebox(0,0)[r]{$\bigl<\bar{\psi}\psi\bigr>$}}
\put(820,-50){\makebox(0,0)[r]{$\beta$}}
\put(176.0,169.0){\rule[-0.200pt]{303.534pt}{0.400pt}}
\put(176.0,68.0){\rule[-0.200pt]{0.400pt}{194.888pt}}
\put(176.0,68.0){\rule[-0.200pt]{4.818pt}{0.400pt}}
\put(154,68){\makebox(0,0)[r]{-0.1}}
\put(1416.0,68.0){\rule[-0.200pt]{4.818pt}{0.400pt}}
\put(176.0,169.0){\rule[-0.200pt]{4.818pt}{0.400pt}}
\put(154,169){\makebox(0,0)[r]{0}}
\put(1416.0,169.0){\rule[-0.200pt]{4.818pt}{0.400pt}}
\put(176.0,270.0){\rule[-0.200pt]{4.818pt}{0.400pt}}
\put(154,270){\makebox(0,0)[r]{0.1}}
\put(1416.0,270.0){\rule[-0.200pt]{4.818pt}{0.400pt}}
\put(176.0,371.0){\rule[-0.200pt]{4.818pt}{0.400pt}}
\put(154,371){\makebox(0,0)[r]{0.2}}
\put(1416.0,371.0){\rule[-0.200pt]{4.818pt}{0.400pt}}
\put(176.0,473.0){\rule[-0.200pt]{4.818pt}{0.400pt}}
\put(154,473){\makebox(0,0)[r]{0.3}}
\put(1416.0,473.0){\rule[-0.200pt]{4.818pt}{0.400pt}}
\put(176.0,574.0){\rule[-0.200pt]{4.818pt}{0.400pt}}
\put(154,574){\makebox(0,0)[r]{0.4}}
\put(1416.0,574.0){\rule[-0.200pt]{4.818pt}{0.400pt}}
\put(176.0,675.0){\rule[-0.200pt]{4.818pt}{0.400pt}}
\put(154,675){\makebox(0,0)[r]{0.5}}
\put(1416.0,675.0){\rule[-0.200pt]{4.818pt}{0.400pt}}
\put(176.0,776.0){\rule[-0.200pt]{4.818pt}{0.400pt}}
\put(154,776){\makebox(0,0)[r]{0.6}}
\put(1416.0,776.0){\rule[-0.200pt]{4.818pt}{0.400pt}}
\put(176.0,877.0){\rule[-0.200pt]{4.818pt}{0.400pt}}
\put(154,877){\makebox(0,0)[r]{0.7}}
\put(1416.0,877.0){\rule[-0.200pt]{4.818pt}{0.400pt}}
\put(176.0,68.0){\rule[-0.200pt]{0.400pt}{4.818pt}}
\put(176,23){\makebox(0,0){0}}
\put(176.0,857.0){\rule[-0.200pt]{0.400pt}{4.818pt}}
\put(309.0,68.0){\rule[-0.200pt]{0.400pt}{4.818pt}}
\put(309,23){\makebox(0,0){0.2}}
\put(309.0,857.0){\rule[-0.200pt]{0.400pt}{4.818pt}}
\put(441.0,68.0){\rule[-0.200pt]{0.400pt}{4.818pt}}
\put(441,23){\makebox(0,0){0.4}}
\put(441.0,857.0){\rule[-0.200pt]{0.400pt}{4.818pt}}
\put(574.0,68.0){\rule[-0.200pt]{0.400pt}{4.818pt}}
\put(574,23){\makebox(0,0){0.6}}
\put(574.0,857.0){\rule[-0.200pt]{0.400pt}{4.818pt}}
\put(707.0,68.0){\rule[-0.200pt]{0.400pt}{4.818pt}}
\put(707,23){\makebox(0,0){0.8}}
\put(707.0,857.0){\rule[-0.200pt]{0.400pt}{4.818pt}}
\put(839.0,68.0){\rule[-0.200pt]{0.400pt}{4.818pt}}
\put(839,23){\makebox(0,0){1}}
\put(839.0,857.0){\rule[-0.200pt]{0.400pt}{4.818pt}}
\put(972.0,68.0){\rule[-0.200pt]{0.400pt}{4.818pt}}
\put(972,23){\makebox(0,0){1.2}}
\put(972.0,857.0){\rule[-0.200pt]{0.400pt}{4.818pt}}
\put(1104.0,68.0){\rule[-0.200pt]{0.400pt}{4.818pt}}
\put(1104,23){\makebox(0,0){1.4}}
\put(1104.0,857.0){\rule[-0.200pt]{0.400pt}{4.818pt}}
\put(1237.0,68.0){\rule[-0.200pt]{0.400pt}{4.818pt}}
\put(1237,23){\makebox(0,0){1.6}}
\put(1237.0,857.0){\rule[-0.200pt]{0.400pt}{4.818pt}}
\put(1370.0,68.0){\rule[-0.200pt]{0.400pt}{4.818pt}}
\put(1370,23){\makebox(0,0){1.8}}
\put(1370.0,857.0){\rule[-0.200pt]{0.400pt}{4.818pt}}
\put(176.0,68.0){\rule[-0.200pt]{303.534pt}{0.400pt}}
\put(1436.0,68.0){\rule[-0.200pt]{0.400pt}{194.888pt}}
\put(176.0,877.0){\rule[-0.200pt]{303.534pt}{0.400pt}}
\put(176.0,68.0){\rule[-0.200pt]{0.400pt}{194.888pt}}
\put(342,793){\raisebox{-.8pt}{\makebox(0,0){$\Diamond$}}}
\put(508,764){\raisebox{-.8pt}{\makebox(0,0){$\Diamond$}}}
\put(673,634){\raisebox{-.8pt}{\makebox(0,0){$\Diamond$}}}
\put(707,568){\raisebox{-.8pt}{\makebox(0,0){$\Diamond$}}}
\put(773,448){\raisebox{-.8pt}{\makebox(0,0){$\Diamond$}}}
\put(839,313){\raisebox{-.8pt}{\makebox(0,0){$\Diamond$}}}
\put(905,229){\raisebox{-.8pt}{\makebox(0,0){$\Diamond$}}}
\put(1005,191){\raisebox{-.8pt}{\makebox(0,0){$\Diamond$}}}
\put(1038,183){\raisebox{-.8pt}{\makebox(0,0){$\Diamond$}}}
\put(1104,176){\raisebox{-.8pt}{\makebox(0,0){$\Diamond$}}}
\put(1171,173){\raisebox{-.8pt}{\makebox(0,0){$\Diamond$}}}
\put(1337,173){\raisebox{-.8pt}{\makebox(0,0){$\Diamond$}}}
\put(342.0,790.0){\rule[-0.200pt]{0.400pt}{1.445pt}}
\put(332.0,790.0){\rule[-0.200pt]{4.818pt}{0.400pt}}
\put(332.0,796.0){\rule[-0.200pt]{4.818pt}{0.400pt}}
\put(508.0,762.0){\rule[-0.200pt]{0.400pt}{1.204pt}}
\put(498.0,762.0){\rule[-0.200pt]{4.818pt}{0.400pt}}
\put(498.0,767.0){\rule[-0.200pt]{4.818pt}{0.400pt}}
\put(673.0,632.0){\rule[-0.200pt]{0.400pt}{1.204pt}}
\put(663.0,632.0){\rule[-0.200pt]{4.818pt}{0.400pt}}
\put(663.0,637.0){\rule[-0.200pt]{4.818pt}{0.400pt}}
\put(707.0,566.0){\rule[-0.200pt]{0.400pt}{1.204pt}}
\put(697.0,566.0){\rule[-0.200pt]{4.818pt}{0.400pt}}
\put(697.0,571.0){\rule[-0.200pt]{4.818pt}{0.400pt}}
\put(773.0,446.0){\rule[-0.200pt]{0.400pt}{0.964pt}}
\put(763.0,446.0){\rule[-0.200pt]{4.818pt}{0.400pt}}
\put(763.0,450.0){\rule[-0.200pt]{4.818pt}{0.400pt}}
\put(839.0,312.0){\rule[-0.200pt]{0.400pt}{0.723pt}}
\put(829.0,312.0){\rule[-0.200pt]{4.818pt}{0.400pt}}
\put(829.0,315.0){\rule[-0.200pt]{4.818pt}{0.400pt}}
\put(905.0,228.0){\rule[-0.200pt]{0.400pt}{0.482pt}}
\put(895.0,228.0){\rule[-0.200pt]{4.818pt}{0.400pt}}
\put(895.0,230.0){\rule[-0.200pt]{4.818pt}{0.400pt}}
\put(1005.0,190.0){\rule[-0.200pt]{0.400pt}{0.482pt}}
\put(995.0,190.0){\rule[-0.200pt]{4.818pt}{0.400pt}}
\put(995.0,192.0){\rule[-0.200pt]{4.818pt}{0.400pt}}
\put(1038.0,182.0){\rule[-0.200pt]{0.400pt}{0.482pt}}
\put(1028.0,182.0){\rule[-0.200pt]{4.818pt}{0.400pt}}
\put(1028.0,184.0){\rule[-0.200pt]{4.818pt}{0.400pt}}
\put(1104.0,176.0){\usebox{\plotpoint}}
\put(1094.0,176.0){\rule[-0.200pt]{4.818pt}{0.400pt}}
\put(1094.0,177.0){\rule[-0.200pt]{4.818pt}{0.400pt}}
\put(1171.0,172.0){\rule[-0.200pt]{0.400pt}{0.482pt}}
\put(1161.0,172.0){\rule[-0.200pt]{4.818pt}{0.400pt}}
\put(1161.0,174.0){\rule[-0.200pt]{4.818pt}{0.400pt}}
\put(1337.0,172.0){\rule[-0.200pt]{0.400pt}{0.482pt}}
\put(1327.0,172.0){\rule[-0.200pt]{4.818pt}{0.400pt}}
\put(1327.0,174.0){\rule[-0.200pt]{4.818pt}{0.400pt}}
\put(229,664){\makebox(0,0){$+$}}
\put(242,671){\makebox(0,0){$+$}}
\put(309,641){\makebox(0,0){$+$}}
\put(342,606){\makebox(0,0){$+$}}
\put(375,526){\makebox(0,0){$+$}}
\put(408,370){\makebox(0,0){$+$}}
\put(441,221){\makebox(0,0){$+$}}
\put(474,184){\makebox(0,0){$+$}}
\put(508,177){\makebox(0,0){$+$}}
\put(541,174){\makebox(0,0){$+$}}
\put(673,171){\makebox(0,0){$+$}}
\put(839,170){\makebox(0,0){$+$}}
\put(1005,170){\makebox(0,0){$+$}}
\put(1171,169){\makebox(0,0){$+$}}

\put(219.00,661.00){\rule[-0.200pt]{4.818pt}{0.400pt}}
\put(219.00,666.00){\rule[-0.200pt]{4.818pt}{0.400pt}}

\put(232.00,668.00){\rule[-0.200pt]{4.818pt}{0.400pt}}
\put(232.00,673.00){\rule[-0.200pt]{4.818pt}{0.400pt}}

\put(299.00,638.00){\rule[-0.200pt]{4.818pt}{0.400pt}}
\put(299.00,643.00){\rule[-0.200pt]{4.818pt}{0.400pt}}

\put(332.00,604.00){\rule[-0.200pt]{4.818pt}{0.400pt}}
\put(332.00,609.00){\rule[-0.200pt]{4.818pt}{0.400pt}}

\put(365.00,524.00){\rule[-0.200pt]{4.818pt}{0.400pt}}
\put(365.00,529.00){\rule[-0.200pt]{4.818pt}{0.400pt}}

\put(398.00,368.00){\rule[-0.200pt]{4.818pt}{0.400pt}}
\put(398.00,372.00){\rule[-0.200pt]{4.818pt}{0.400pt}}

\put(431.00,219.00){\rule[-0.200pt]{4.818pt}{0.400pt}}
\put(431.00,222.00){\rule[-0.200pt]{4.818pt}{0.400pt}}

\put(464.00,183.00){\rule[-0.200pt]{4.818pt}{0.400pt}}
\put(464.00,185.00){\rule[-0.200pt]{4.818pt}{0.400pt}}

\put(498.00,177.00){\rule[-0.200pt]{4.818pt}{0.400pt}}
\put(498.00,178.00){\rule[-0.200pt]{4.818pt}{0.400pt}}

\put(531.00,173.00){\rule[-0.200pt]{4.818pt}{0.400pt}}
\put(531.00,174.00){\rule[-0.200pt]{4.818pt}{0.400pt}}

\put(663.00,170.00){\rule[-0.200pt]{4.818pt}{0.400pt}}
\put(663.00,171.00){\rule[-0.200pt]{4.818pt}{0.400pt}}

\put(829.00,169.00){\rule[-0.200pt]{4.818pt}{0.400pt}}
\put(829.00,170.00){\rule[-0.200pt]{4.818pt}{0.400pt}}

\put(995.00,169.00){\rule[-0.200pt]{4.818pt}{0.400pt}}
\put(995.00,170.00){\rule[-0.200pt]{4.818pt}{0.400pt}}

\put(1161.00,168.00){\rule[-0.200pt]{4.818pt}{0.400pt}}
\put(1161.00,169.00){\rule[-0.200pt]{4.818pt}{0.400pt}}

\sbox{\plotpoint}{\rule[-0.400pt]{0.800pt}{0.800pt}}%
\put(229,507){\raisebox{-.8pt}{\makebox(0,0){$\Box$}}}
\put(242,498){\raisebox{-.8pt}{\makebox(0,0){$\Box$}}}
\put(275,448){\raisebox{-.8pt}{\makebox(0,0){$\Box$}}}
\put(282,407){\raisebox{-.8pt}{\makebox(0,0){$\Box$}}}
\put(289,365){\raisebox{-.8pt}{\makebox(0,0){$\Box$}}}
\put(295,306){\raisebox{-.8pt}{\makebox(0,0){$\Box$}}}
\put(302,230){\raisebox{-.8pt}{\makebox(0,0){$\Box$}}}
\put(309,187){\raisebox{-.8pt}{\makebox(0,0){$\Box$}}}
\put(322,166){\raisebox{-.8pt}{\makebox(0,0){$\Box$}}}
\put(342,171){\raisebox{-.8pt}{\makebox(0,0){$\Box$}}}
\put(375,172){\raisebox{-.8pt}{\makebox(0,0){$\Box$}}}
\put(441,170){\raisebox{-.8pt}{\makebox(0,0){$\Box$}}}
\put(508,169){\raisebox{-.8pt}{\makebox(0,0){$\Box$}}}
\put(673,170){\raisebox{-.8pt}{\makebox(0,0){$\Box$}}}
\put(839,169){\raisebox{-.8pt}{\makebox(0,0){$\Box$}}}
\put(229.0,504.0){\rule[-0.400pt]{0.800pt}{1.204pt}}
\put(219.0,504.0){\rule[-0.400pt]{4.818pt}{0.800pt}}
\put(219.0,509.0){\rule[-0.400pt]{4.818pt}{0.800pt}}
\put(242.0,495.0){\rule[-0.400pt]{0.800pt}{1.204pt}}
\put(232.0,495.0){\rule[-0.400pt]{4.818pt}{0.800pt}}
\put(232.0,500.0){\rule[-0.400pt]{4.818pt}{0.800pt}}
\put(275.0,446.0){\rule[-0.400pt]{0.800pt}{0.964pt}}
\put(265.0,446.0){\rule[-0.400pt]{4.818pt}{0.800pt}}
\put(265.0,450.0){\rule[-0.400pt]{4.818pt}{0.800pt}}
\put(282.0,405.0){\rule[-0.400pt]{0.800pt}{0.964pt}}
\put(272.0,405.0){\rule[-0.400pt]{4.818pt}{0.800pt}}
\put(272.0,409.0){\rule[-0.400pt]{4.818pt}{0.800pt}}
\put(289.0,363.0){\rule[-0.400pt]{0.800pt}{0.964pt}}
\put(279.0,363.0){\rule[-0.400pt]{4.818pt}{0.800pt}}
\put(279.0,367.0){\rule[-0.400pt]{4.818pt}{0.800pt}}
\put(295.0,304.0){\rule[-0.400pt]{0.800pt}{0.964pt}}
\put(285.0,304.0){\rule[-0.400pt]{4.818pt}{0.800pt}}
\put(285.0,308.0){\rule[-0.400pt]{4.818pt}{0.800pt}}
\put(302.0,229.0){\usebox{\plotpoint}}
\put(292.0,229.0){\rule[-0.400pt]{4.818pt}{0.800pt}}
\put(292.0,232.0){\rule[-0.400pt]{4.818pt}{0.800pt}}
\put(309.0,186.0){\usebox{\plotpoint}}
\put(299.0,186.0){\rule[-0.400pt]{4.818pt}{0.800pt}}
\put(299.0,188.0){\rule[-0.400pt]{4.818pt}{0.800pt}}
\put(322.0,165.0){\usebox{\plotpoint}}
\put(312.0,165.0){\rule[-0.400pt]{4.818pt}{0.800pt}}
\put(312.0,167.0){\rule[-0.400pt]{4.818pt}{0.800pt}}
\put(342.0,170.0){\usebox{\plotpoint}}
\put(332.0,170.0){\rule[-0.400pt]{4.818pt}{0.800pt}}
\put(332.0,171.0){\rule[-0.400pt]{4.818pt}{0.800pt}}
\put(375.0,171.0){\usebox{\plotpoint}}
\put(365.0,171.0){\rule[-0.400pt]{4.818pt}{0.800pt}}
\put(365.0,173.0){\rule[-0.400pt]{4.818pt}{0.800pt}}
\put(441.0,169.0){\usebox{\plotpoint}}
\put(431.0,169.0){\rule[-0.400pt]{4.818pt}{0.800pt}}
\put(431.0,170.0){\rule[-0.400pt]{4.818pt}{0.800pt}}
\put(508.0,169.0){\usebox{\plotpoint}}
\put(498.0,169.0){\rule[-0.400pt]{4.818pt}{0.800pt}}
\put(498.0,170.0){\rule[-0.400pt]{4.818pt}{0.800pt}}
\put(673.0,170.0){\usebox{\plotpoint}}
\put(663.0,170.0){\rule[-0.400pt]{4.818pt}{0.800pt}}
\put(663.0,171.0){\rule[-0.400pt]{4.818pt}{0.800pt}}
\put(839.0,169.0){\usebox{\plotpoint}}
\put(829.0,169.0){\rule[-0.400pt]{4.818pt}{0.800pt}}
\put(829.0,170.0){\rule[-0.400pt]{4.818pt}{0.800pt}}
\end{picture}
\vspace{4mm}

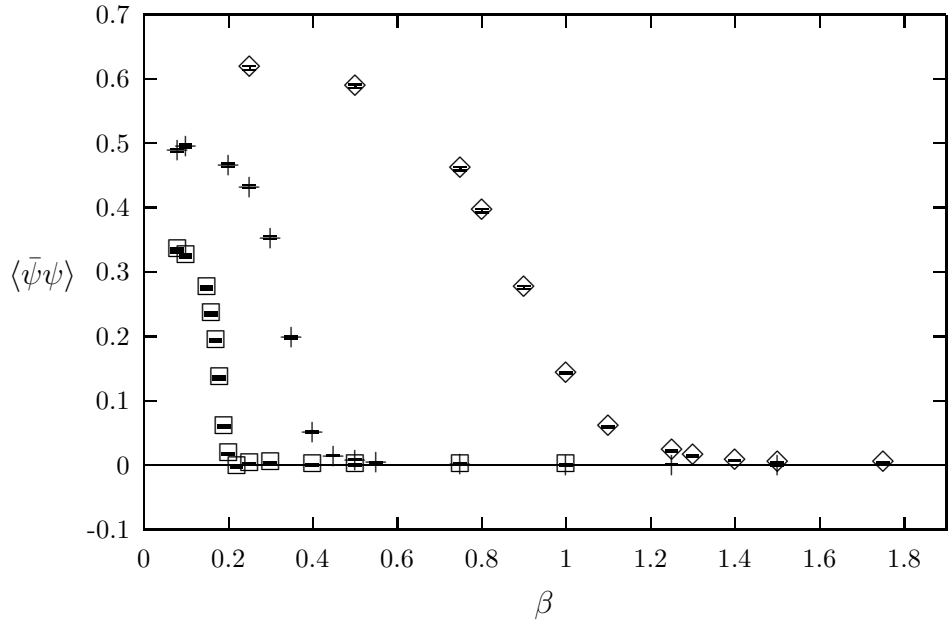
Figure 1: The chiral condensate, $\bigl<\bar{\psi}\psi\bigr>_{m\rightarrow
0}$, vs. $\beta$ 
for $N=2$ ($\Diamond$ points), $N=4$ ($+$ points), and $N=6$ ($\Box$ points)
on a $8^3$ lattice.
\vspace{4mm}

\setlength{\unitlength}{0.240900pt}
\ifx\plotpoint\undefined\newsavebox{\plotpoint}\fi
\sbox{\plotpoint}{\rule[-0.200pt]{0.400pt}{0.400pt}}%
\begin{picture}(1500,900)(0,0)
\font\gnuplot=cmr10 at 10pt
\gnuplot
\sbox{\plotpoint}{\rule[-0.200pt]{0.400pt}{0.400pt}}%
\put(60,468){\makebox(0,0)[r]{$\beta^{2}\bigl<\bar{\psi}\psi\bigr>$}}
\put(820,-50){\makebox(0,0)[r]{$\beta$}}
\put(176.0,68.0){\rule[-0.200pt]{303.534pt}{0.400pt}}
\put(176.0,68.0){\rule[-0.200pt]{4.818pt}{0.400pt}}
\put(154,68){\makebox(0,0)[r]{0}}
\put(1416.0,68.0){\rule[-0.200pt]{4.818pt}{0.400pt}}
\put(176.0,203.0){\rule[-0.200pt]{4.818pt}{0.400pt}}
\put(154,203){\makebox(0,0)[r]{0.05}}
\put(1416.0,203.0){\rule[-0.200pt]{4.818pt}{0.400pt}}
\put(176.0,338.0){\rule[-0.200pt]{4.818pt}{0.400pt}}
\put(154,338){\makebox(0,0)[r]{0.1}}
\put(1416.0,338.0){\rule[-0.200pt]{4.818pt}{0.400pt}}
\put(176.0,472.0){\rule[-0.200pt]{4.818pt}{0.400pt}}
\put(154,472){\makebox(0,0)[r]{0.15}}
\put(1416.0,472.0){\rule[-0.200pt]{4.818pt}{0.400pt}}
\put(176.0,607.0){\rule[-0.200pt]{4.818pt}{0.400pt}}
\put(154,607){\makebox(0,0)[r]{0.2}}
\put(1416.0,607.0){\rule[-0.200pt]{4.818pt}{0.400pt}}
\put(176.0,742.0){\rule[-0.200pt]{4.818pt}{0.400pt}}
\put(154,742){\makebox(0,0)[r]{0.25}}
\put(1416.0,742.0){\rule[-0.200pt]{4.818pt}{0.400pt}}
\put(176.0,877.0){\rule[-0.200pt]{4.818pt}{0.400pt}}
\put(154,877){\makebox(0,0)[r]{0.3}}
\put(1416.0,877.0){\rule[-0.200pt]{4.818pt}{0.400pt}}
\put(291.0,68.0){\rule[-0.200pt]{0.400pt}{4.818pt}}
\put(291,23){\makebox(0,0){0.8}}
\put(291.0,857.0){\rule[-0.200pt]{0.400pt}{4.818pt}}
\put(520.0,68.0){\rule[-0.200pt]{0.400pt}{4.818pt}}
\put(520,23){\makebox(0,0){1}}
\put(520.0,857.0){\rule[-0.200pt]{0.400pt}{4.818pt}}
\put(749.0,68.0){\rule[-0.200pt]{0.400pt}{4.818pt}}
\put(749,23){\makebox(0,0){1.2}}
\put(749.0,857.0){\rule[-0.200pt]{0.400pt}{4.818pt}}
\put(978.0,68.0){\rule[-0.200pt]{0.400pt}{4.818pt}}
\put(978,23){\makebox(0,0){1.4}}
\put(978.0,857.0){\rule[-0.200pt]{0.400pt}{4.818pt}}
\put(1207.0,68.0){\rule[-0.200pt]{0.400pt}{4.818pt}}
\put(1207,23){\makebox(0,0){1.6}}
\put(1207.0,857.0){\rule[-0.200pt]{0.400pt}{4.818pt}}
\put(1436.0,68.0){\rule[-0.200pt]{0.400pt}{4.818pt}}
\put(1436,23){\makebox(0,0){1.8}}
\put(1436.0,857.0){\rule[-0.200pt]{0.400pt}{4.818pt}}
\put(176.0,68.0){\rule[-0.200pt]{303.534pt}{0.400pt}}
\put(1436.0,68.0){\rule[-0.200pt]{0.400pt}{194.888pt}}
\put(176.0,877.0){\rule[-0.200pt]{303.534pt}{0.400pt}}
\put(176.0,68.0){\rule[-0.200pt]{0.400pt}{194.888pt}}
\put(233,766){\raisebox{-.8pt}{\makebox(0,0){$\Diamond$}}}
\put(291,749){\raisebox{-.8pt}{\makebox(0,0){$\Diamond$}}}
\put(405,671){\raisebox{-.8pt}{\makebox(0,0){$\Diamond$}}}
\put(520,452){\raisebox{-.8pt}{\makebox(0,0){$\Diamond$}}}
\put(634,261){\raisebox{-.8pt}{\makebox(0,0){$\Diamond$}}}
\put(806,160){\raisebox{-.8pt}{\makebox(0,0){$\Diamond$}}}
\put(863,129){\raisebox{-.8pt}{\makebox(0,0){$\Diamond$}}}
\put(978,106){\raisebox{-.8pt}{\makebox(0,0){$\Diamond$}}}
\put(1092,90){\raisebox{-.8pt}{\makebox(0,0){$\Diamond$}}}
\put(1379,100){\raisebox{-.8pt}{\makebox(0,0){$\Diamond$}}}
\put(233.0,762.0){\rule[-0.200pt]{0.400pt}{1.686pt}}
\put(223.0,762.0){\rule[-0.200pt]{4.818pt}{0.400pt}}
\put(223.0,769.0){\rule[-0.200pt]{4.818pt}{0.400pt}}
\put(291.0,745.0){\rule[-0.200pt]{0.400pt}{1.927pt}}
\put(281.0,745.0){\rule[-0.200pt]{4.818pt}{0.400pt}}
\put(281.0,753.0){\rule[-0.200pt]{4.818pt}{0.400pt}}
\put(405.0,666.0){\rule[-0.200pt]{0.400pt}{2.168pt}}
\put(395.0,666.0){\rule[-0.200pt]{4.818pt}{0.400pt}}
\put(395.0,675.0){\rule[-0.200pt]{4.818pt}{0.400pt}}
\put(520.0,448.0){\rule[-0.200pt]{0.400pt}{2.168pt}}
\put(510.0,448.0){\rule[-0.200pt]{4.818pt}{0.400pt}}
\put(510.0,457.0){\rule[-0.200pt]{4.818pt}{0.400pt}}
\put(634.0,257.0){\rule[-0.200pt]{0.400pt}{2.168pt}}
\put(624.0,257.0){\rule[-0.200pt]{4.818pt}{0.400pt}}
\put(624.0,266.0){\rule[-0.200pt]{4.818pt}{0.400pt}}
\put(806.0,156.0){\rule[-0.200pt]{0.400pt}{1.927pt}}
\put(796.0,156.0){\rule[-0.200pt]{4.818pt}{0.400pt}}
\put(796.0,164.0){\rule[-0.200pt]{4.818pt}{0.400pt}}
\put(863.0,125.0){\rule[-0.200pt]{0.400pt}{1.927pt}}
\put(853.0,125.0){\rule[-0.200pt]{4.818pt}{0.400pt}}
\put(853.0,133.0){\rule[-0.200pt]{4.818pt}{0.400pt}}
\put(978.0,102.0){\rule[-0.200pt]{0.400pt}{1.927pt}}
\put(968.0,102.0){\rule[-0.200pt]{4.818pt}{0.400pt}}
\put(968.0,110.0){\rule[-0.200pt]{4.818pt}{0.400pt}}
\put(1092.0,86.0){\rule[-0.200pt]{0.400pt}{1.927pt}}
\put(1082.0,86.0){\rule[-0.200pt]{4.818pt}{0.400pt}}
\put(1082.0,94.0){\rule[-0.200pt]{4.818pt}{0.400pt}}
\put(1379.0,95.0){\rule[-0.200pt]{0.400pt}{2.168pt}}
\put(1369.0,95.0){\rule[-0.200pt]{4.818pt}{0.400pt}}
\put(1369.0,104.0){\rule[-0.200pt]{4.818pt}{0.400pt}}
\end{picture}
\vspace{4mm}

Figure 2: $\beta^{2}\bigl<\bar{\psi}\psi\bigr>_{m\rightarrow 0}$ vs. $\beta$
for $N=2$ on a $8^3$ lattice.

To elucidate these behaviors
more clearly, in Fig. 2, we search for a continuum scaling window 
for $N=2$. Like QED3 $N=0$ and $1$ case, the chiral condensate exhibits a long 
plateau in weak coupling region. On the other hand, $N=4$ shows two slopes
and hints flat region.

\vspace{7mm}

\setlength{\unitlength}{0.240900pt}
\ifx\plotpoint\undefined\newsavebox{\plotpoint}\fi
\begin{picture}(1500,900)(0,0)
\font\gnuplot=cmr10 at 10pt
\gnuplot
\sbox{\plotpoint}{\rule[-0.200pt]{0.400pt}{0.400pt}}%
\put(70,468){\makebox(0,0)[r]{$\bigl<\bar{\psi}\psi\bigr>^{2}$}}
\put(820,-50){\makebox(0,0)[r]{$\beta$}}
\put(176.0,158.0){\rule[-0.200pt]{303.534pt}{0.400pt}}
\put(176.0,68.0){\rule[-0.200pt]{0.400pt}{194.888pt}}
\put(176.0,68.0){\rule[-0.200pt]{4.818pt}{0.400pt}}
\put(154,68){\makebox(0,0)[r]{-0.05}}
\put(1416.0,68.0){\rule[-0.200pt]{4.818pt}{0.400pt}}
\put(176.0,158.0){\rule[-0.200pt]{4.818pt}{0.400pt}}
\put(154,158){\makebox(0,0)[r]{0}}
\put(1416.0,158.0){\rule[-0.200pt]{4.818pt}{0.400pt}}
\put(176.0,248.0){\rule[-0.200pt]{4.818pt}{0.400pt}}
\put(154,248){\makebox(0,0)[r]{0.05}}
\put(1416.0,248.0){\rule[-0.200pt]{4.818pt}{0.400pt}}
\put(176.0,338.0){\rule[-0.200pt]{4.818pt}{0.400pt}}
\put(154,338){\makebox(0,0)[r]{0.1}}
\put(1416.0,338.0){\rule[-0.200pt]{4.818pt}{0.400pt}}
\put(176.0,428.0){\rule[-0.200pt]{4.818pt}{0.400pt}}
\put(154,428){\makebox(0,0)[r]{0.15}}
\put(1416.0,428.0){\rule[-0.200pt]{4.818pt}{0.400pt}}
\put(176.0,517.0){\rule[-0.200pt]{4.818pt}{0.400pt}}
\put(154,517){\makebox(0,0)[r]{0.2}}
\put(1416.0,517.0){\rule[-0.200pt]{4.818pt}{0.400pt}}
\put(176.0,607.0){\rule[-0.200pt]{4.818pt}{0.400pt}}
\put(154,607){\makebox(0,0)[r]{0.25}}
\put(1416.0,607.0){\rule[-0.200pt]{4.818pt}{0.400pt}}
\put(176.0,697.0){\rule[-0.200pt]{4.818pt}{0.400pt}}
\put(154,697){\makebox(0,0)[r]{0.3}}
\put(1416.0,697.0){\rule[-0.200pt]{4.818pt}{0.400pt}}
\put(176.0,787.0){\rule[-0.200pt]{4.818pt}{0.400pt}}
\put(154,787){\makebox(0,0)[r]{0.35}}
\put(1416.0,787.0){\rule[-0.200pt]{4.818pt}{0.400pt}}
\put(176.0,877.0){\rule[-0.200pt]{4.818pt}{0.400pt}}
\put(154,877){\makebox(0,0)[r]{0.4}}
\put(1416.0,877.0){\rule[-0.200pt]{4.818pt}{0.400pt}}
\put(176.0,68.0){\rule[-0.200pt]{0.400pt}{4.818pt}}
\put(176,23){\makebox(0,0){0}}
\put(176.0,857.0){\rule[-0.200pt]{0.400pt}{4.818pt}}
\put(316.0,68.0){\rule[-0.200pt]{0.400pt}{4.818pt}}
\put(316,23){\makebox(0,0){0.2}}
\put(316.0,857.0){\rule[-0.200pt]{0.400pt}{4.818pt}}
\put(456.0,68.0){\rule[-0.200pt]{0.400pt}{4.818pt}}
\put(456,23){\makebox(0,0){0.4}}
\put(456.0,857.0){\rule[-0.200pt]{0.400pt}{4.818pt}}
\put(596.0,68.0){\rule[-0.200pt]{0.400pt}{4.818pt}}
\put(596,23){\makebox(0,0){0.6}}
\put(596.0,857.0){\rule[-0.200pt]{0.400pt}{4.818pt}}
\put(736.0,68.0){\rule[-0.200pt]{0.400pt}{4.818pt}}
\put(736,23){\makebox(0,0){0.8}}
\put(736.0,857.0){\rule[-0.200pt]{0.400pt}{4.818pt}}
\put(876.0,68.0){\rule[-0.200pt]{0.400pt}{4.818pt}}
\put(876,23){\makebox(0,0){1}}
\put(876.0,857.0){\rule[-0.200pt]{0.400pt}{4.818pt}}
\put(1016.0,68.0){\rule[-0.200pt]{0.400pt}{4.818pt}}
\put(1016,23){\makebox(0,0){1.2}}
\put(1016.0,857.0){\rule[-0.200pt]{0.400pt}{4.818pt}}
\put(1156.0,68.0){\rule[-0.200pt]{0.400pt}{4.818pt}}
\put(1156,23){\makebox(0,0){1.4}}
\put(1156.0,857.0){\rule[-0.200pt]{0.400pt}{4.818pt}}
\put(1296.0,68.0){\rule[-0.200pt]{0.400pt}{4.818pt}}
\put(1296,23){\makebox(0,0){1.6}}
\put(1296.0,857.0){\rule[-0.200pt]{0.400pt}{4.818pt}}
\put(1436.0,68.0){\rule[-0.200pt]{0.400pt}{4.818pt}}
\put(1436,23){\makebox(0,0){1.8}}
\put(1436.0,857.0){\rule[-0.200pt]{0.400pt}{4.818pt}}
\put(176.0,68.0){\rule[-0.200pt]{303.534pt}{0.400pt}}
\put(1436.0,68.0){\rule[-0.200pt]{0.400pt}{194.888pt}}
\put(176.0,877.0){\rule[-0.200pt]{303.534pt}{0.400pt}}
\put(176.0,68.0){\rule[-0.200pt]{0.400pt}{194.888pt}}
\put(351,842){\raisebox{-.8pt}{\makebox(0,0){$\Diamond$}}}
\put(526,781){\raisebox{-.8pt}{\makebox(0,0){$\Diamond$}}}
\put(701,538){\raisebox{-.8pt}{\makebox(0,0){$\Diamond$}}}
\put(736,438){\raisebox{-.8pt}{\makebox(0,0){$\Diamond$}}}
\put(806,295){\raisebox{-.8pt}{\makebox(0,0){$\Diamond$}}}
\put(876,194){\raisebox{-.8pt}{\makebox(0,0){$\Diamond$}}}
\put(946,164){\raisebox{-.8pt}{\makebox(0,0){$\Diamond$}}}
\put(1051,159){\raisebox{-.8pt}{\makebox(0,0){$\Diamond$}}}
\put(1086,158){\raisebox{-.8pt}{\makebox(0,0){$\Diamond$}}}
\put(1156,158){\raisebox{-.8pt}{\makebox(0,0){$\Diamond$}}}
\put(1226,158){\raisebox{-.8pt}{\makebox(0,0){$\Diamond$}}}
\put(1401,158){\raisebox{-.8pt}{\makebox(0,0){$\Diamond$}}}
\put(351.0,836.0){\rule[-0.200pt]{0.400pt}{2.891pt}}
\put(341.0,836.0){\rule[-0.200pt]{4.818pt}{0.400pt}}
\put(341.0,848.0){\rule[-0.200pt]{4.818pt}{0.400pt}}
\put(526.0,775.0){\rule[-0.200pt]{0.400pt}{2.891pt}}
\put(516.0,775.0){\rule[-0.200pt]{4.818pt}{0.400pt}}
\put(516.0,787.0){\rule[-0.200pt]{4.818pt}{0.400pt}}
\put(701.0,534.0){\rule[-0.200pt]{0.400pt}{1.927pt}}
\put(691.0,534.0){\rule[-0.200pt]{4.818pt}{0.400pt}}
\put(691.0,542.0){\rule[-0.200pt]{4.818pt}{0.400pt}}
\put(736.0,435.0){\rule[-0.200pt]{0.400pt}{1.445pt}}
\put(726.0,435.0){\rule[-0.200pt]{4.818pt}{0.400pt}}
\put(726.0,441.0){\rule[-0.200pt]{4.818pt}{0.400pt}}
\put(806.0,293.0){\rule[-0.200pt]{0.400pt}{0.964pt}}
\put(796.0,293.0){\rule[-0.200pt]{4.818pt}{0.400pt}}
\put(796.0,297.0){\rule[-0.200pt]{4.818pt}{0.400pt}}
\put(876.0,194.0){\usebox{\plotpoint}}
\put(866.0,194.0){\rule[-0.200pt]{4.818pt}{0.400pt}}
\put(866.0,195.0){\rule[-0.200pt]{4.818pt}{0.400pt}}
\put(946,164){\usebox{\plotpoint}}
\put(936.0,164.0){\rule[-0.200pt]{4.818pt}{0.400pt}}
\put(936.0,164.0){\rule[-0.200pt]{4.818pt}{0.400pt}}
\put(1051,159){\usebox{\plotpoint}}
\put(1041.0,159.0){\rule[-0.200pt]{4.818pt}{0.400pt}}
\put(1041.0,159.0){\rule[-0.200pt]{4.818pt}{0.400pt}}
\put(1086,158){\usebox{\plotpoint}}
\put(1076.0,158.0){\rule[-0.200pt]{4.818pt}{0.400pt}}
\put(1076.0,158.0){\rule[-0.200pt]{4.818pt}{0.400pt}}
\put(1156,158){\usebox{\plotpoint}}
\put(1146.0,158.0){\rule[-0.200pt]{4.818pt}{0.400pt}}
\put(1146.0,158.0){\rule[-0.200pt]{4.818pt}{0.400pt}}
\put(1226,158){\usebox{\plotpoint}}
\put(1216.0,158.0){\rule[-0.200pt]{4.818pt}{0.400pt}}
\put(1216.0,158.0){\rule[-0.200pt]{4.818pt}{0.400pt}}
\put(1401,158){\usebox{\plotpoint}}
\put(1391.0,158.0){\rule[-0.200pt]{4.818pt}{0.400pt}}
\put(1391.0,158.0){\rule[-0.200pt]{4.818pt}{0.400pt}}
\put(232,588){\makebox(0,0){$+$}}
\put(246,600){\makebox(0,0){$+$}}
\put(316,549){\makebox(0,0){$+$}}
\put(351,494){\makebox(0,0){$+$}}
\put(386,382){\makebox(0,0){$+$}}
\put(421,229){\makebox(0,0){$+$}}
\put(456,163){\makebox(0,0){$+$}}
\put(491,158){\makebox(0,0){$+$}}
\put(526,158){\makebox(0,0){$+$}}
\put(561,158){\makebox(0,0){$+$}}
\put(701,158){\makebox(0,0){$+$}}
\put(876,158){\makebox(0,0){$+$}}
\put(1051,158){\makebox(0,0){$+$}}
\put(1226,158){\makebox(0,0){$+$}}

\put(222.00,583.00){\rule[-0.200pt]{4.818pt}{0.400pt}}
\put(222.00,592.00){\rule[-0.200pt]{4.818pt}{0.400pt}}

\put(236.00,595.00){\rule[-0.200pt]{4.818pt}{0.400pt}}
\put(236.00,605.00){\rule[-0.200pt]{4.818pt}{0.400pt}}

\put(306.00,544.00){\rule[-0.200pt]{4.818pt}{0.400pt}}
\put(306.00,553.00){\rule[-0.200pt]{4.818pt}{0.400pt}}

\put(341.00,490.00){\rule[-0.200pt]{4.818pt}{0.400pt}}
\put(341.00,498.00){\rule[-0.200pt]{4.818pt}{0.400pt}}

\put(376.00,380.00){\rule[-0.200pt]{4.818pt}{0.400pt}}
\put(376.00,385.00){\rule[-0.200pt]{4.818pt}{0.400pt}}

\put(411.00,227.00){\rule[-0.200pt]{4.818pt}{0.400pt}}
\put(411.00,230.00){\rule[-0.200pt]{4.818pt}{0.400pt}}

\put(446.00,162.00){\rule[-0.200pt]{4.818pt}{0.400pt}}
\put(446.00,163.00){\rule[-0.200pt]{4.818pt}{0.400pt}}

\put(481.00,158.00){\rule[-0.200pt]{4.818pt}{0.400pt}}
\put(481.00,158.00){\rule[-0.200pt]{4.818pt}{0.400pt}}

\put(516.00,158.00){\rule[-0.200pt]{4.818pt}{0.400pt}}
\put(516.00,158.00){\rule[-0.200pt]{4.818pt}{0.400pt}}

\put(551.00,158.00){\rule[-0.200pt]{4.818pt}{0.400pt}}
\put(551.00,158.00){\rule[-0.200pt]{4.818pt}{0.400pt}}

\put(691.00,158.00){\rule[-0.200pt]{4.818pt}{0.400pt}}
\put(691.00,158.00){\rule[-0.200pt]{4.818pt}{0.400pt}}

\put(866.00,158.00){\rule[-0.200pt]{4.818pt}{0.400pt}}
\put(866.00,158.00){\rule[-0.200pt]{4.818pt}{0.400pt}}

\put(1041.00,158.00){\rule[-0.200pt]{4.818pt}{0.400pt}}
\put(1041.00,158.00){\rule[-0.200pt]{4.818pt}{0.400pt}}

\put(1216.00,158.00){\rule[-0.200pt]{4.818pt}{0.400pt}}
\put(1216.00,158.00){\rule[-0.200pt]{4.818pt}{0.400pt}}

\sbox{\plotpoint}{\rule[-0.400pt]{0.800pt}{0.800pt}}%
\put(232,358){\raisebox{-.8pt}{\makebox(0,0){$\Box$}}}
\put(246,347){\raisebox{-.8pt}{\makebox(0,0){$\Box$}}}
\put(281,294){\raisebox{-.8pt}{\makebox(0,0){$\Box$}}}
\put(288,257){\raisebox{-.8pt}{\makebox(0,0){$\Box$}}}
\put(295,225){\raisebox{-.8pt}{\makebox(0,0){$\Box$}}}
\put(302,191){\raisebox{-.8pt}{\makebox(0,0){$\Box$}}}
\put(309,164){\raisebox{-.8pt}{\makebox(0,0){$\Box$}}}
\put(316,158){\raisebox{-.8pt}{\makebox(0,0){$\Box$}}}
\put(330,158){\raisebox{-.8pt}{\makebox(0,0){$\Box$}}}
\put(351,158){\raisebox{-.8pt}{\makebox(0,0){$\Box$}}}
\put(386,158){\raisebox{-.8pt}{\makebox(0,0){$\Box$}}}
\put(456,158){\raisebox{-.8pt}{\makebox(0,0){$\Box$}}}
\put(526,158){\raisebox{-.8pt}{\makebox(0,0){$\Box$}}}
\put(701,158){\raisebox{-.8pt}{\makebox(0,0){$\Box$}}}
\put(876,158){\raisebox{-.8pt}{\makebox(0,0){$\Box$}}}
\put(232.0,356.0){\rule[-0.400pt]{0.800pt}{1.204pt}}
\put(222.0,356.0){\rule[-0.400pt]{4.818pt}{0.800pt}}
\put(222.0,361.0){\rule[-0.400pt]{4.818pt}{0.800pt}}
\put(246.0,345.0){\rule[-0.400pt]{0.800pt}{1.204pt}}
\put(236.0,345.0){\rule[-0.400pt]{4.818pt}{0.800pt}}
\put(236.0,350.0){\rule[-0.400pt]{4.818pt}{0.800pt}}
\put(281.0,292.0){\rule[-0.400pt]{0.800pt}{0.964pt}}
\put(271.0,292.0){\rule[-0.400pt]{4.818pt}{0.800pt}}
\put(271.0,296.0){\rule[-0.400pt]{4.818pt}{0.800pt}}
\put(288.0,255.0){\rule[-0.400pt]{0.800pt}{0.964pt}}
\put(278.0,255.0){\rule[-0.400pt]{4.818pt}{0.800pt}}
\put(278.0,259.0){\rule[-0.400pt]{4.818pt}{0.800pt}}
\put(295.0,224.0){\usebox{\plotpoint}}
\put(285.0,224.0){\rule[-0.400pt]{4.818pt}{0.800pt}}
\put(285.0,227.0){\rule[-0.400pt]{4.818pt}{0.800pt}}
\put(302.0,190.0){\usebox{\plotpoint}}
\put(292.0,190.0){\rule[-0.400pt]{4.818pt}{0.800pt}}
\put(292.0,192.0){\rule[-0.400pt]{4.818pt}{0.800pt}}
\put(309.0,164.0){\usebox{\plotpoint}}
\put(299.0,164.0){\rule[-0.400pt]{4.818pt}{0.800pt}}
\put(299.0,165.0){\rule[-0.400pt]{4.818pt}{0.800pt}}
\put(316.0,158.0){\usebox{\plotpoint}}
\put(306.0,158.0){\rule[-0.400pt]{4.818pt}{0.800pt}}
\put(306.0,159.0){\rule[-0.400pt]{4.818pt}{0.800pt}}
\put(330,158){\usebox{\plotpoint}}
\put(320.0,158.0){\rule[-0.400pt]{4.818pt}{0.800pt}}
\put(320.0,158.0){\rule[-0.400pt]{4.818pt}{0.800pt}}
\put(351,158){\usebox{\plotpoint}}
\put(341.0,158.0){\rule[-0.400pt]{4.818pt}{0.800pt}}
\put(341.0,158.0){\rule[-0.400pt]{4.818pt}{0.800pt}}
\put(386,158){\usebox{\plotpoint}}
\put(376.0,158.0){\rule[-0.400pt]{4.818pt}{0.800pt}}
\put(376.0,158.0){\rule[-0.400pt]{4.818pt}{0.800pt}}
\put(456,158){\usebox{\plotpoint}}
\put(446.0,158.0){\rule[-0.400pt]{4.818pt}{0.800pt}}
\put(446.0,158.0){\rule[-0.400pt]{4.818pt}{0.800pt}}
\put(526,158){\usebox{\plotpoint}}
\put(516.0,158.0){\rule[-0.400pt]{4.818pt}{0.800pt}}
\put(516.0,158.0){\rule[-0.400pt]{4.818pt}{0.800pt}}
\put(701,158){\usebox{\plotpoint}}
\put(691.0,158.0){\rule[-0.400pt]{4.818pt}{0.800pt}}
\put(691.0,158.0){\rule[-0.400pt]{4.818pt}{0.800pt}}
\put(876,158){\usebox{\plotpoint}}
\put(866.0,158.0){\rule[-0.400pt]{4.818pt}{0.800pt}}
\put(866.0,158.0){\rule[-0.400pt]{4.818pt}{0.800pt}}
\end{picture}
\vspace{6mm}

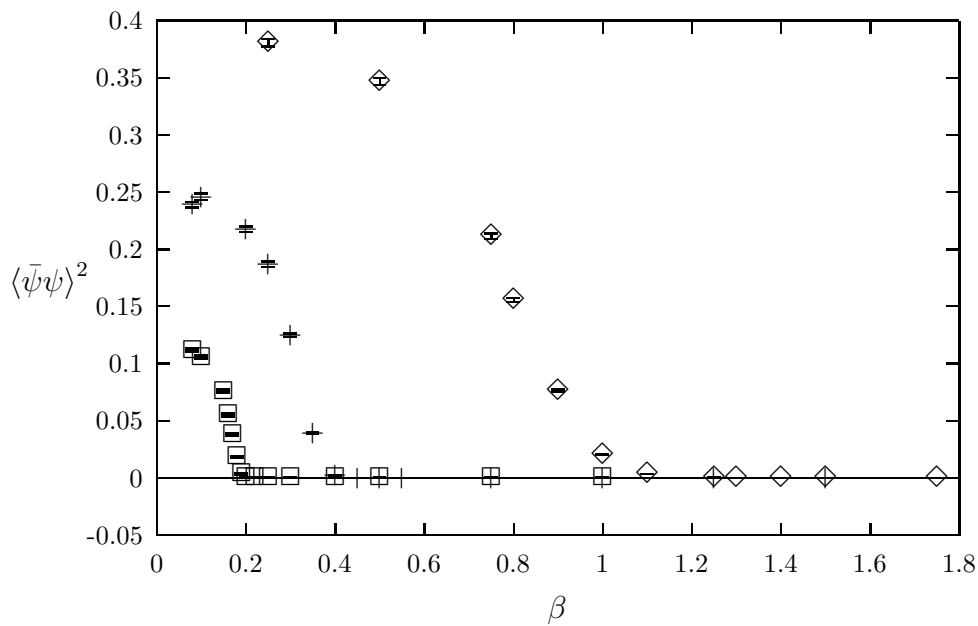
Figure 3: $\bigl<\bar{\psi}\psi\bigr>^2_{m\rightarrow 0}$ vs. $\beta$
for $N=2$($\Diamond$), $N=4$($+$), and $N=6$($\Box$) on a $8^3$ lattice.
\vspace{6mm}

In Fig. 3, we plot $\bigl<\bar{\psi}\psi\bigr>^2_{m\rightarrow 0}$ vs. $\beta$,
in order to see whether the critical behavior follows a mean field theory,
\begin{equation}
\bigl<\bar{\psi}\psi\bigr> \sim (\beta_c - \beta)^\frac{1}{2}.
\end{equation}
It appears that $N = 2$ data does not agree with a mean field behavior 
and $N = 6$ data seems to follow a mean field expectation.
For $N = 4$ data, it is very hard to say whether the system behaves in one way
or the other. From internal energy density and specific heat peak, we obtain 
$\beta_c = 0.9(1)$ for $N = 2$, $0.35(5)$ for $N = 4$, and 
$0.18(1)$ for $N = 6$. These values are roughly consistent with those obtained
from Fig. 1, 2, and 3.

According to the QED3 case, large finite volume effect is expected 
\cite{DKK,HK}. In order to purchase this effect, simulations on larger 
lattices and smaller $m$ values are needed. Thus, we are investigating the 
same system on a $16^3$ lattice with the masses $m = 0.05, 0.025$ and $0.0125$.
In Table 1, we compare $\bigl<\bar{\psi}\psi\bigr>_{m\rightarrow 0}$ for $N=2$ 
and in Table 2 compare that for $N=6$ on $8^4$ and $16^3$ volume. The first 
column is the chiral condensate for $8^3$ using a linear extrapolation of 
all range of $m$ data, the second column is that for $16^3$ using a linear 
extrapolation of smaller $m$ data ($m = 0.025,\;0.0125$), and the third
column is that for $16^3$ using a quadratic fit to existing data. 
Although the dependence of the chiral condensate on both the size of 
lattices and the extrapolation method are considerable quantitatively
(particularly in the very weak coupling region of $N=2$ case), 
we found that the overall qualitative nature of the chiral condensate does 
not seem to be changed: In all three cases, $N=6$ data are compatible with 
mean-field behavior but $N=2$ data is not. 

\vspace{2mm}

\begin{center}
\begin{tabular}{|c|c|c|c|}
\hline
$\hspace{7mm}\beta\hspace{7mm}$&$\bigl<\bar{\psi}\psi\bigr> (8^3)$
		&$\bigl<\bar{\psi}\psi\bigr> (16^3)$ 
		&$\bigl<\bar{\psi}\psi\bigr>_{quad} (16^3)$ \\
\hline
0.5    & 0.589(3) & 0.641(9) & 0.643  \\
0.75   & 0.460(2) & 0.493(8) & 0.488  \\
1.0    & 0.142(2) & 0.207(4) & 0.195  \\
1.25   & 0.0218(9)& 0.053(2) & 0.048  \\
1.3    & 0.0134(8)& 0.019(2) & 0.007  \\
1.4    & 0.0071(7)& 0.003(2) &-0.009  \\
\hline
\end{tabular}

\vspace{4mm}
Table 1: Comparison of the chiral condensate for $N=2$

\vspace{5mm}

\begin{tabular}{|c|c|c|c|}
\hline
$\hspace{7mm}\beta\hspace{7mm}$ &$\bigl<\bar{\psi}\psi\bigr> (8^3)$
		&$\bigl<\bar{\psi}\psi\bigr> (16^3)$ 
		&$\bigl<\bar{\psi}\psi\bigr>_{quad} (16^3)$ \\
\hline
0.18   & 0.135(2) & 0.263(4) & 0.239  \\
0.2    & 0.017(1) & 0.085(2) & 0.058  \\
0.22   &-0.004(1) & 0.031(2) & 0.017  \\
0.25   & 0.0014(8)& 0.004(9) &-0.003  \\ 
0.3    & 0.0027(7)&-0.028(1) &-0.010  \\
0.35   &          & 0.004(1) & 0.003  \\
0.4    & 0.0004(6)& 0.0004(8)&-0.003  \\
0.45   &          & 0.0006(10)&-0.001 \\
0.5    & 0.0002(5)& 0.0005(7)&-0.001  \\
\hline
\end{tabular}

\vspace{4mm}
Table 2: Comparison of the chiral condensate for $N=6$
\end{center}

In conclusion, we have found spontaneous chiral symmetry breaking 
for $N = 2, \;4$ and 6 in 3D Thirring model by use of Monte Carlo 
simulation on a $8^3$ and a $16^3$ lattice. However, the chiral condensate in 
the $m \rightarrow 0$ limit suggests that the nature of the phase 
transition in $N = 2$ is different from that in $N = 6$. $N = 2$ case 
has a long tail of non-vanishing chiral condensate in weak coupling region
and $N = 6$ case has a phase separation between the strong coupling
region and the weak coupling region. It suggests a possibility that there
exists $N_{\rm cr}$ where the chiral phase transition becomes
a mean field type transition. This situation is quite similar to
QED3. In contrast to the lattice simulation result of the auxiliary vector 
field formulation, our result seems to be more consistent with the gauge 
theory formulation of 3D Thirring model in continuum. 

For a future work, we think 
that further detailed study on finite volume effects is needed 
although we have discussed our preliminary data on a $16^3$ lattice.
Also, two further questions remain open: One is whether lattice 3D Thirring 
model is similar to 3D lattice QED. The other is whether the induced gauge 
field kinetic term takes the form of compact QED-like or noncompact QED-like.
The latter issue will be determined by the importance of topological sector.
At this moment, we do not know what the induced term looks like.

\vspace{7mm}

We would like to thank Ph. de Forcrand, D.K. Hong, J.B. Kogut, P. Maris, 
D.K. Sinclair, H.S. Song
and K. Yamawaki for various comments and discussions. S. Ohta and the 
Computing Center of the Institute of Physical and Chemical 
Research (RIKEN) are acknowledged for their support and allowing us 
to access their Fujitsu VPP-500. 
This work was supported by the KOSEF(through CTP,
Brain Pool Program and 95-0702-04-01-3) and the Korean Ministry of Education
(BSRI-95-2413).

\def\hebibliography#1{\begin{center}\subsection*{References
}\end{center}\list
  {[\arabic{enumi}]}{\settowidth\labelwidth{[#1]}
\leftmargin\labelwidth	  \advance\leftmargin\labelsep
    \usecounter{enumi}}
    \def\newblock{\hskip .11em plus .33em minus .07em}
    \sloppy\clubpenalty4000\widowpenalty4000
    \sfcode`\.=1000\relax}

\let\endhebibliography=\endlist

\begin{hebibliography}{100}
\bibitem{Par} G. Parisi, Nucl. Phys. B {\bf 100}, 368 (1975); S. Hikami 
and T. Muta, Prog. Theor. Phys. {\bf 57}, 785 (1977); Z. Yang, Texas 
preprint UTTG-40-90 (unpublished).
\bibitem{KK} N.V. Krasnikov and A.B. Kyatkin, Mod. Phys. Lett. A {\bf 6}, 
1315 (1991).
\bibitem{Han} S. Hands, Phys. Rev. D {\bf 51}, 5816 (1995). 
\bibitem{GMRS} M. Gomes, R.S. Mendes, R.F. Ribeiro and A.J. da Silva, 
Phys. Rev. D {\bf 43}, 3516 (1991).
\bibitem{HP} D.K. Hong and S.H. Park, Phys. Rev. D {\bf 49}, 5507 (1994).
\bibitem{HLY} S.J. Hyun, G.H. Lee and J.H. Yee, Phys. Rev. D {\bf 50},
6542 (1994); Y.M. Ahn, B.K. Chung, J.-M. Chung and Q-H. Park, Kyung Hee Preprint 
KHTP-94-05 (unpublished).
\bibitem{RS} G. Rossini and F.A. Schaposnik, Phys. Lett. B {\bf 338},
465 (1994).
\bibitem{IKSY} T. Itoh, M. Sugiura, Y. Kim and K. Yamawaki, Prog. Theor. 
Phys. {\bf 93}, 417 (1995).
\bibitem{Kon} K.-I. Kondo, Nucl. Phys. B {\bf 450}, 251 (1995). 
\bibitem{DH} L.D. Debbio and S. Hands, Preprint SWAT/95/93, hep-lat/9512013.
\bibitem{BKY} M. Bando, T. Kugo and K. Yamawaki, Phys. Rep. {\bf 164},
217 (1988).
\bibitem{VW} With $N$ four-component fermions in (2+1)D, parity-violating
mass $\bigl<\bar{\psi}i\gamma^{0}\gamma^{1}\gamma^{2}\psi\bigr>$ can be generated
dynamically, however, in this gauge invariant formulation
of the Thirring model, such symmetry breaking is proven to be energetically
unfavorable by using exact argument in C. Vafa and E. Witten, Commun. Math.
Phys. {\bf 95}, 257 (1984) (see Ref.\cite{IKSY}). 
\bibitem{Red} Similarly, only under the guardianship of gauge symmetry,
we can clarify this issue as shown in \cite{IKSY}: One can take the 
regularization to keep
both $U(1)$ gauge symmetry and parity without any ambiguity for the
four Dirac-fermion case (equivalent to even-number two-component 
Dirac-fermions), however the Chern-Simons term identified as the 
parity-violating anomaly should arise in odd-number two-component 
Dirac-fermion case as in 3D quantum electrodynamics (QED3).
\bibitem{SSY} A. Shibata, M. Sugiura and K. Yamawaki, in preparation.
\bibitem{BB} C. Burden and A.N. Burkitt, Europhys. Lett., {\bf 3}, 545 (1987).
\bibitem{Kim_Kim} S. Kim and Y. Kim, work in progress.
\bibitem{DKK} E. Dagotto, A. Koci\'c and J.B. Kogut, Phys. Rev. 
Lett.  {\bf 62}, 1083 (1989); E. Dagotto, J.B. Kogut and A. Koci\'c, 
Nucl. Phys. B {\bf 334}, 279 (1990).
\bibitem{ANW} T.W. Appelquist, D. Nash and L.C.R. Wijewardhana, Phys. Rev.
Lett. {\bf 60}, 2575 (1988); T. Matsuki, L. Miao and K.S. Viswanathan, Simon
Fraser University preprint; For a review, see C.D. Roberts and A.G. Williams,
Prog. Part. Nucl. Phys. {\bf 33}, 477 (1994).
\bibitem{HK} S. Hands and J.B. Kogut, Nucl. Phys. B {\bf 335}, 455 (1990).
\end{hebibliography}
\end{document}